# Audio enabled information extraction system for cricket and hockey domains

S.Saraswathi, Narasimha Sravan.V, Sai Vamsi Krishna.B.V and Suresh Reddy.S

**Abstract-** The proposed system aims at the retrieval of the summarized information from the documents collected from web based search engine as per the user query related to cricket and hockey domain. The system is designed in a manner that it takes the voice commands as keywords for search. The parts of speech in the query are extracted using the natural language extractor for English. Based on the keywords the search is categorized into 2 types: - 1.Concept wise – information retrieved to the query is retrieved based on the keywords and the concept words related to it. The retrieved information is summarized using the probabilistic approach and weighted means algorithm.2.Keyword search – extracts the result relevant to the query from the highly ranked document retrieved from the search by the search engine. The relevant search results are retrieved and then keywords are used for summarizing part. During summarization it follows the weighted and probabilistic approaches in order to identify the data comparable to the keywords extracted. The extracted information is then refined repeatedly through the aggregation process to reduce redundancy. Finally the resultant data is submitted to the user in the form of audio output.

**Index Terms**—Tagger, Summarization, Information extraction, Information retrieval, Keyword search, Concept-wise search.

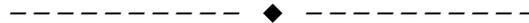

## 1. INTRODUCTION

Information Extraction is the process of retrieving relevant information from the available data on the basis of the keywords provided. In this process first the input is obtained from the user in the form of speech query and is converted into text [1]. The obtained text is processed by a Parts Of Speech tagger [2] which detects the keywords for deciding the type of search. This leads to Concept wise search or the Keyword search based on the 'wh' keywords obtained [3].

During the process of extraction, it is dynamically decided to select the components thereby optimizing the summarization ratio. This proposal has two important phases, information retrieval and summarization. We propose a new approach for this based on the following studies.

Clustering algorithms [4] such as K-Means have popularly been used as semantic summarization methods where cluster centers become the summarized set. The goal of semantic summarization is to provide a summarized view of the original dataset such that the summarization ratio is maximized while the error (i.e., information loss) is minimized.
In Discovery Net [5], each distributed dataset is locally summarized by the K-Means algorithm. Then, the summarized sets are sent to a central site for global clustering. The quality of this approach is largely dependent on the performance of K-means.

A scalable clustering algorithm [6] is proposed to deal with very large datasets. In this approach, the datasets are divided into several equally sized and disjoint segments. Then, hard K-Means or Fuzzy K-Means algorithm are used to summarize each data segment. Similar to Discovery Net, a clustering algorithm is then run on the union of summarized sets.

A database system consists of millions of data items that could be picked out as their priorities by proper approach. At first, all the data is classified into different categories. Each category is assigned with a predefined priority. The higher the priority information has more possibility to be chosen. Secondly, each data can only be visited once. In addition, writing a program to perform the task can be very straightforward [7]. However, it is not very easy to design an algorithm that is most efficient for all scenarios.

The technique used to search keyword query dynamically generates new pages, called composed pages [8], which contain all query keywords. The composed pages are generated by extracting and stitching together relevant pieces from hyper-linked Web pages, and retaining links to the original Web pages. To rank the composed pages, the authors consider both the hyper link structure of the original pages, as well as the associations between the keywords within each page.

The rest of this paper is organized as follows. In Section 2, a detailed explanation of the proposed system is discussed. In Section 3, the drawbacks of

————————————————

The authors acknowledge All India Council of Technical education (AICTE) which has funded under Research Promotion Scheme for carrying out this project.
- *F.A. Author is Associate Professor, in the Department of Information Technology, Pondicherry Engineering College, Pondicherry, India.*
- *S.B. Author are Under graduate students in the Department of Information Technology, Pondicherry Engineering College, Pondicherry, India.*

the existing system and the betterments made in the proposed system are discussed. In Section 4, testing parameters for the system have been described. Section 5 is the conclusion.

## 2. PROPOSED WORK

*MODULE DESCRIPTION AND DESIGN*:
The entire system is classified such that it consists of the following major divisions as reffered from figure 1

1. Input query in speech format.
2. Conversion of speech to text.
3. Extraction of keywords
4. Categorization of query
    4.1. Keyword search identification.
    4.2. Concept-wise search identification.
5. Keyword search
    5.1. Document extraction.
    5.2. Information retrieval.
6. Concept-wise search
    6.1. Relevant document extraction.
    6.2. Information extraction from resultant documents.
        6.2.1. Weighted means algorithm.
        6.2.2. Probabilistic approach.
    6.3. Aggregation of the resultant information.
7. Output
    7.1. Conversion of text to speech.

### 1. Input query

The input to the system is a voice Command, Query, any keyword in cricket or hockey domains which is to be in general english language. The queries may be related to instances like – get score, information regarding stadium, players or about the series and regarding the biographies of the players.

### 2. Conversion to text

The obtained query in the form of speech is then passed to Dragan natural speaking software. Dragon natural speaking (v 9.10)[8] is used for taking user voice commands as input and to convert them to text. System requirements are: CPU of 2.4 GHz with dual core processor, a minimum of 512 MB RAM, hard disk space of 1 GB, L2 Cache of minimum 512 KB. Most accurate (20% more compared to previous version), fast, vast vocabulary are the attractive traits that proved valuable for the proposed system from this tool.

### 3. Keyword Extraction

Part-of-speech tagging (POS tagging or POST), also called grammatical tagging or word-category disambiguation, is the process of marking up the words in a text (corpus) as corresponding to a particular part of speech, based on both its definition, as well as its context —i.e. relationship with adjacent and related words in a phrase, sentence, or paragraph. Dynamic programming algorithms are applied for tagging in POS tagger [9]. The obtained query from the dragon natural speaking software, in the text format, is operated with POS tagger in order to extract the nouns and the pronouns in the given query.The needed information is to be searched in the world wide web inorder to obtain the information relevent to the query in the form of audio output after the summarization is carried out.

### 4. Categorization of query

The query from the user can be classified broadly as
- Key word search query
- Conceptwise search query

#### 4.1. Keyword search identification
The obatained results in the POS tagger output is verified for the keywords 'Which', 'Where', 'What', 'Who', 'When'. If the match is found, then the query is a generic one and then is classified as the Keyword search category. For example consider the queries
a) What is the score of sachin tendulakar in world cup 2007?
Ans:1796
b) When is the next hockey world cup?
Ans:2014

#### 4.2. Concept-wise search identification
If the query is related to a particular person or any of the mentioned categories the entire information related to the queries will be grouped based on the predetermined concepts for the particular query. To group the related data tree structure[10] is used for better and efficient search reasults as shown in the figure 2. This increases both the vastness of the subject covered by the process and also the quality[11] of the retrieved documents as all the aspects of the related domains are addressed.
For example consider the queries
a) Dhanraj pillai
b) Childhood of Dhyan chand

### 5. Keyword search

#### 5.1. Document extraction
The obtained keywords from the POS Tagger are passed to the web browser in order to obtain the web results for the relevant query. From the obtained results, the most relevant result is extracted and is converted to the text format using an *html to text converter*.

#### 5.2. Information retrieval
The document that is obtained as the result of the text converter, based on the keywords available in the query is subjected to match with the keywords found in the extracted document. Then the most relevent answer is found based on the number of matches and the obtained result is then passed to the text to speech converter.

### 6. Concept-wise search

#### 6.1. Document extraction

The obtained keywords from the POS Tagger are passed on to a pre-built tree structure[11][12] inorder to obtain the relevant concept words for carrying out the required concept-wise search as shown in figure 2. The obtained concept words are then passed on to the web browser to search extensively on the various aspects of the keyword that has been given in the query. The top most relevant web results are obtained and are converted to the text documents using the *html to text converter*.

The sample run information extraction for the given domains has been depicted in the figure 3.

#### 6.2. Information extraction from resultant documents

The obtained text documents are then subjected to two algorithms individually and sequentially inorder to obatin the relevant information from each document. The two algorithmic appraoches are explained below.

#### 6.2.1 Weighted means algorithm:

Each of the text document is divided into the components based on the strategy described in the weighted as well as the probability based algorithms.In the weight based algorithm, a binomial distribution function is applied over the set of lines or the component chosen and the probability of the component is checked and compared for extraction. Similar to that of the probability based algorithm, weight assigned to the noun and pronoun play the decisive role in extracting the components. The algorithms are applied over the document and the required component is extracted from the document. The same procedure is applied over the remaining documents inorder to obtain the components related to the query.

The keywords obtained after applying Pos tagger on the query are stored in a keyword array. The keywords in the array are passed to the concept tree diagram as shown in figure 2, in order to obtain the related concept words. These keywords are updated to the keyword array. Two variables N and Pn are intialised to note the weight of the nouns and the pronouns in the components of the documents respectively. In addition a result array is initialised inorder to store values of the weigths of the componets in the documents while execution of the algorithm.

ALGORITHM I

Step 1: Fetch one of the resultant documents in the text format.
Step 2: Divide the entire document into components of size ten lines each.
Step 3: Fetch the first line of the document and check for all the keywords from the keyword array.
Step 4: If the keyword is a noun and there is match increment the noun counter.
Step 5: If the keyword is a pronoun and there is match increment the pronoun counter.
Step 6: Repeat step 3 to 5 until all the lines in the document are exhausted.
Step 7: Evaluate the weight of the component based on the formula below

$$w = 2 * N + Pn \qquad (1)$$

step 8: Store the weight of the component in the result array.
Step 9: Repeat from step 3 to 8 for all the rest of the components in the document.
Step 10: Compare the values of the result array of weights and fetch the first three components as per the result array sort.
Step 11: Repeat step 2 to 10 for all the other extracted documents.
Step 12: Store all the extracted components from all the documents to the resultant document for the aggregation.

Each search on a keyword/concept word gives a result that can be segregated as a concept named under that keyword/concept word.

The above algorithm gives a final resultant document that consists of extracted information from all the web results. This document is passed for the probabilistic algorithm.

#### 6.3. Aggregation

#### 6.3.1 Probabilistic approach

The resultant components obtained from each of the documents are applied with probabilistic algorithm as below.

The POS tagger is applied to the first line of the document and the results are stored in a temporary array. Then the subsequent lines are applied with tagger and the keywords are extracted. The number of keywords in first line (n1) are matched with the number of keywords in the subsequent lines one by one (n2).

- If n1 == n2 AND all the keywords match OR nearly 3/4$^{th}$ keywords match, then discard the later line

- If n1>n2 AND n2 is a subset of n1 or vice versa then discard the later line

Once the related components are obtained from the results, the aggregation process is carried out on the resultant document. Aggregation is carried out such that the redundancy of the information in the various extarcted components are weighed against each other and redundant data will be removed. Once the aggregation is completed the final document is passed to the output section. This is because of the possibility of similar data extracted from multiple sources.

Initially the number of lines in the document is assessed and stored in a temporary variable.

### ALGORITHM II

Step 1: fetch the resultant document/ component.
Step 2: fetch the first line and apply the POs tagger to it.
Step 3: the obtained results of the tagger are stored in the temporary arrays for nouns and pronouns separately.
Step 4: fetch Next line and apply the POS tagger and fetch the results.
Step 5: compare the results of step 3 and step 4 to find a match.
Step 6: increment the success pointer P in case of the match.
Step 7: once the document is exhausted, evaluate the probability distribution of the keywords, PRO, of the line as below

$$PRO = jC_p * \left(\frac{P}{10}\right)^P * \left(1 - \left(\frac{P}{10}\right)\right)^{(4-P)} \quad (2)$$

Where j stands for the jth line in the iteration.
Step 8: if the value is more than threshold value, 0.5, discard the first line.
Step 9: else retain the first line.
Step 10: repeat from step 1 to 9 with next line of the document.

The above algorithm checks the threshold value is found as the average of the probability of each document retrieved. This also checks the redundancy of the data in the resultant document. The data based on the above algorithm will remove the redundancy.

## 7. Output

### 7.1. Conversion of text to speech

The aggregated document that contains the refined information required by the user will be converted to the speech format by the text to speech converter.

Voice XML Editor (IVR) is the Text to Speech converter that has been used. When user develops any new project for IVR system he uses a lot of phrases that help him to navigate through IVR scripit. These phrases often require several changes during development process. It is very convenient and almost impossible to record new phrases every time when changes in script logic are required. Geode IVR software allows user to easily create new prompts (sound files) using available TTS server and incorporate them into IVR script. Recently started free Text-To-Speech service allows any user in any part of the world to use very powerful and convenient TTS feature of our IVR software.

This tool can run independent of Operating System and results are generated in standard UK English in both male and female voices with adjustable speed and volumes.

## 3. COMPARISON

The major modules that are considered in this paper are information extraction and Information retrieval of the web extracted data. The previous algorithms developed for each of these fields are considered and the proposed work is developed based on them, overcoming their drawbacks.

The following is the comparative study between the previous and the proposed works based on modules explained above.

### A. *Heuristic Approach Of Algorithms:*

Most of the existing algorithms use heuristic functionalities without recalculation for the extraction and the summarization process. For example in the weighted information extraction algorithm $\sum_{i}^{p} F_i^* m$ which is a heuristic summation function which fails quality summarization. Because of this there is the problem of information loss and the degradation of the summarization ratio value. In order to overcome that and generate an efficient information extraction algorithm, two stages of summarization, initially weighted summarization (1) and later probabilistic summarization (2). Web algorithm [9] follow the search and the extraction patterns based on the keywords alone thereby leaving the data related to the keywords indirectly reducing the quality of the retrieval. In the proposed algorithm, before the search, the length of the keywords and the search data are compared thus increasing the efficiency. In addition to the keywords the algorithms also considers the related words like pronouns and meta data for information extraction.

### B. *Redundancy In Multiple Documents Extraction*

Existing algorithms, weighted information extraction algorithm [7] and web algorithm [9] can be applied over a single text document for the information extraction to function efficiently and effectively. But they fail utterly when the same algorithm is applied in the multiple document extraction scenarios i.e. a set of information components are retrieved independently from multiple documents and is finally retrieved as a single document. This, in general, with the existing algorithms causes redundancy of the information. In the proposed work, this drawback is overcome by summarizing the data in two steps by two different algorithms repeatedly. The probabilistic approach in the aggregation module, verifies the redundancy and averts it based on the calculative moves.

### 4. PARAMETERS
#### 1) MOS(Mean Opinion Score)

Mean of the remarks obtained from the users of the system.

$$MOS = (OS_1 + OS_2 + OS_3 + \cdots + OS_n)/n \quad (3)$$

Where OS is Opinion Score.

MOS and OS range from 0 to 10.

*2) Summarization Ratio*

It is the ratio of the size of the Summarized text to the size of the original document.

$$Summarization\ ratio = (number\ of\ lines\ in\ summarized\ text) / (number\ of\ lines\ in\ original\ document) \quad (4)$$

Value ranges from 0 to 1.
The above one is calculated for each single document. Effective summarization ratio is calculated for multiple documents by taking average of individual documents.

*3) Precision*

No of relevant documents retrieved divided by the total no of documents retrieved by a search.

$$precision = (number\ of\ relevant\ documents\ retrieved) / (total\ number\ of\ documents\ retrieved) \quad (5)$$

It ranges from 0 to 1.

## 5. RESULTS

### A. KEYWORD SEARCH RESULTS

TABLE 1
KEYWORD SEARCH RESULTS BASED ON QUERY TYPE

| Query Category | No. of Queries in Category | Mean Doc. Extraction Rate | Mean Info. Retrieval Rate | Mean Latency Time(Min) | Fault Count | Mean Opinion Score(10) |
|---|---|---|---|---|---|---|
| Score | 117 | 66.67 | 42 | 0.0425 | 11 | 7.8 |
| Date | 115 | 59.88 | 50 | 0.0367 | 13 | 5.9 |
| Player/Team | 110 | 61 | 49 | 0.0418 | 29 | 8.4 |
| Place | 120 | 61 | 46 | 0.02 | 19 | 7.9 |

**MEAN DOC. EXTRACTION RATE:** Average rate of the documents extracted by the system and given by

$$(\sum_{i=1}^{n}(1/doc.\ Extraction\ latency))/n \quad (6)$$

Where n is the number of documents considered.

**MEAN INFO. RETRIEVAL RATE**: Average rate of the information retrieved by the system and is given by

$$(\sum_{i=1}^{n}(1/Info.retrieval\ latency))/n \quad (7)$$

**MEAN RETRIEVAL LATENCY:** Average of the total latency obtained from the summation of the doc extraction and information retrieval latencies. It is in minutes and is given by

$$\sum_{i=1}^{n}(doc.retrieval\ latency + info.retrieval\ latency)/n \quad (8)$$

**FAULT COUNT:** This is the number of documents that did not project any kind of relevant results.

### B. CONCEPT-WISE SEARCH RESULTS

No. of queries tested: 520
Queries on players/ personalities: 370
Queries on pitches/grounds: 150
Results abstracted effectively on players/ personalities: 341
Results abstracted effectively on pitches/grounds: 137
Precision on players/ personalities: 0.9216
Precision on pitches/grounds: 0.9133
Average no. of web results obtained for each query: 10
Weight assigned for keywords/concept words that are nouns: 1
Weight assigned for pronouns of the keywords /concept words in query: 0.5
No. of components extracted from a document in the weighted means approach:

*components with weight > average weight of the components of the document.*

Threshold value for probabilistic approach: 0.5

More analytical results of concept-wise search are shown below in the tables 2 and 3.

TABLE 2
RESULTS OF QUERIES BASED ON PERSONALITIES

| Queries on personality categories | No. of queries | Results obtained queries | precision | Avg. no. of components per document | Average number of documents used in information retrieval | Avg. no. lines in the resultant document per concept | Summarization ratio of the system on query category |
|---|---|---|---|---|---|---|---|
| Personal | 115 | 108 | 0.9391 | 9.5 | 9.5 | 9 | 0.32 |
| career | 125 | 120 | 0.96 | 9.8 | 9.4 | 12 | 0.35 |
| achievements | 130 | 125 | 0.9615 | 9.8 | 9.5 | 6 | 0.42 |

TABLE 3.
RESULTS OF QUERIES BASED ON GROUND

| Queries on Ground Categories | No. of queries | Results obtained queries | precision | Avg. no. of components per document | Average number of documents used in information retrieval | Avg. no. lines in the resultant document per concept | Summarization ratio of the system on query category |
|---|---|---|---|---|---|---|---|
| Demography | 56 | 52 | 0.9285 | 9.2 | 9.12 | 8 | 0.28 |
| Matches | 94 | 89 | 0.9468 | 8.95 | 8.95 | 5 | 0.317 |

## 6. CONCLUSION

Audio enable information retrieval is a versatile system that can create a notable change in the information browsing wherein ordinary user's thirst for information, which is generally large, to be made as crispy as possible with as informative as possible. In addition the audio enabled schema makes the system user friendly through audio input and audio output.

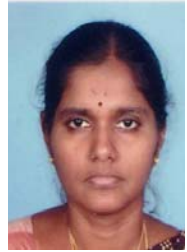

**First A. Author Saraswathi** is Assistant professor, in the Department of Information Technology, Pondicherry Engineering College, Pondicherry, India. She completed her PhD, in the area of speech recognition for Tamil language. Her areas of interest include speech processing, artificial intelligence and expert systems.

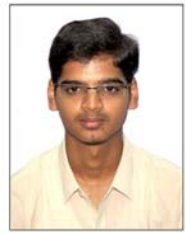

**Second B. Author Jr Narasimha Sravan V** is final year student of Department of Information Technology, Pondicherry Engineering College, Pondicherry, India. His areas of interest are Artificial Intelligence, Robotics.

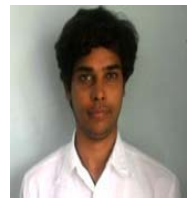

**Second B. Author Jr Sai Vamsi Krishna B V is** final year student of Department of Information Technology, Pondicherry Engineering College, Pondicherry, India. His areas of interest are Artificial Intelligence, Object oriented Programming concepts.

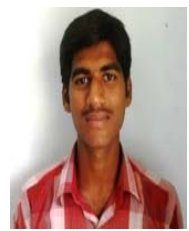

**Second B. Author Jr Suresh Reddy S is** final year student of Department of Information Technology, Pondicherry Engineering College, Pondicherry, India. His areas of interest are Artificial Intelligence, Data Structures.

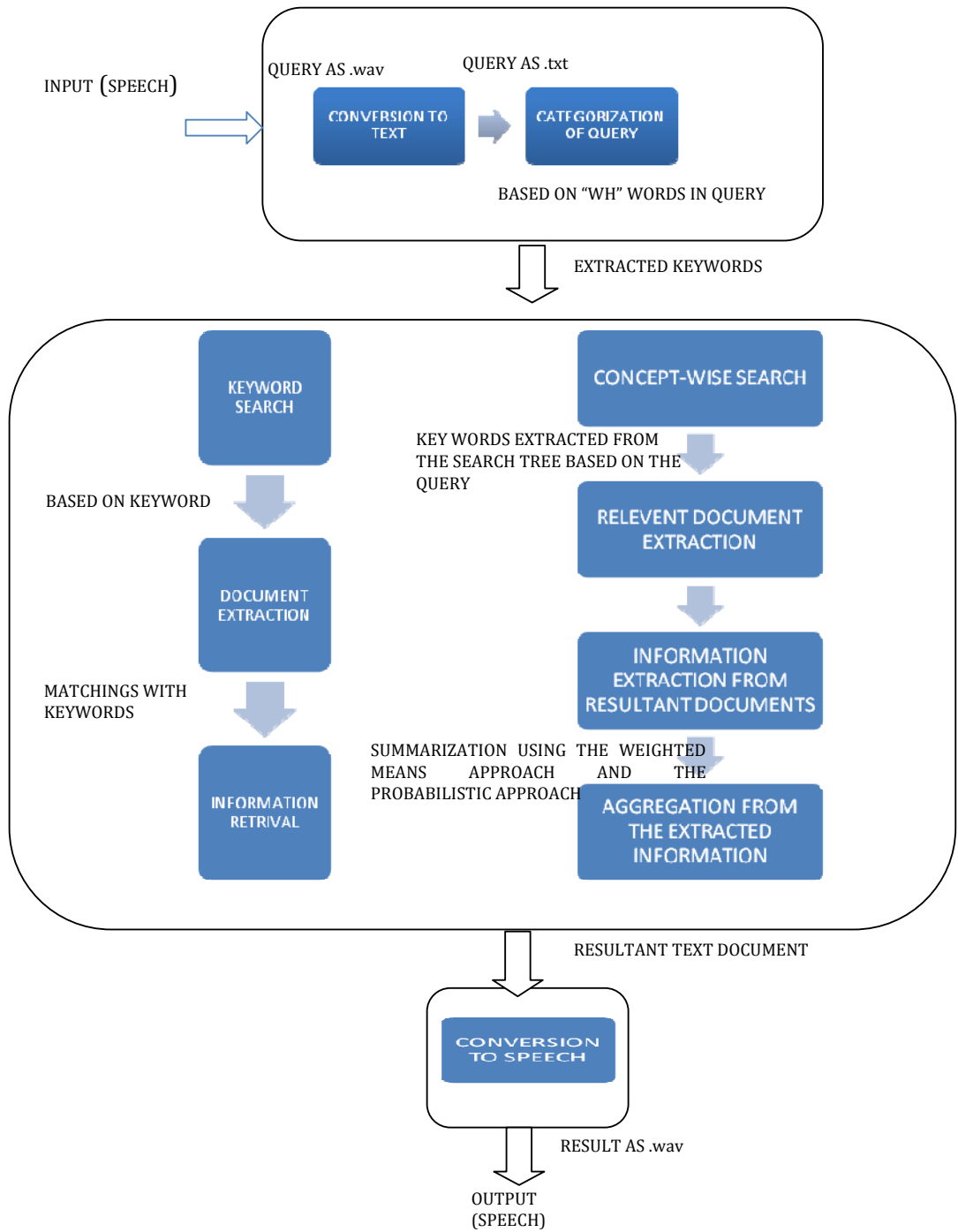

Fig. 1. Over all flow diagram of Audio Enabled Information Retrieval System

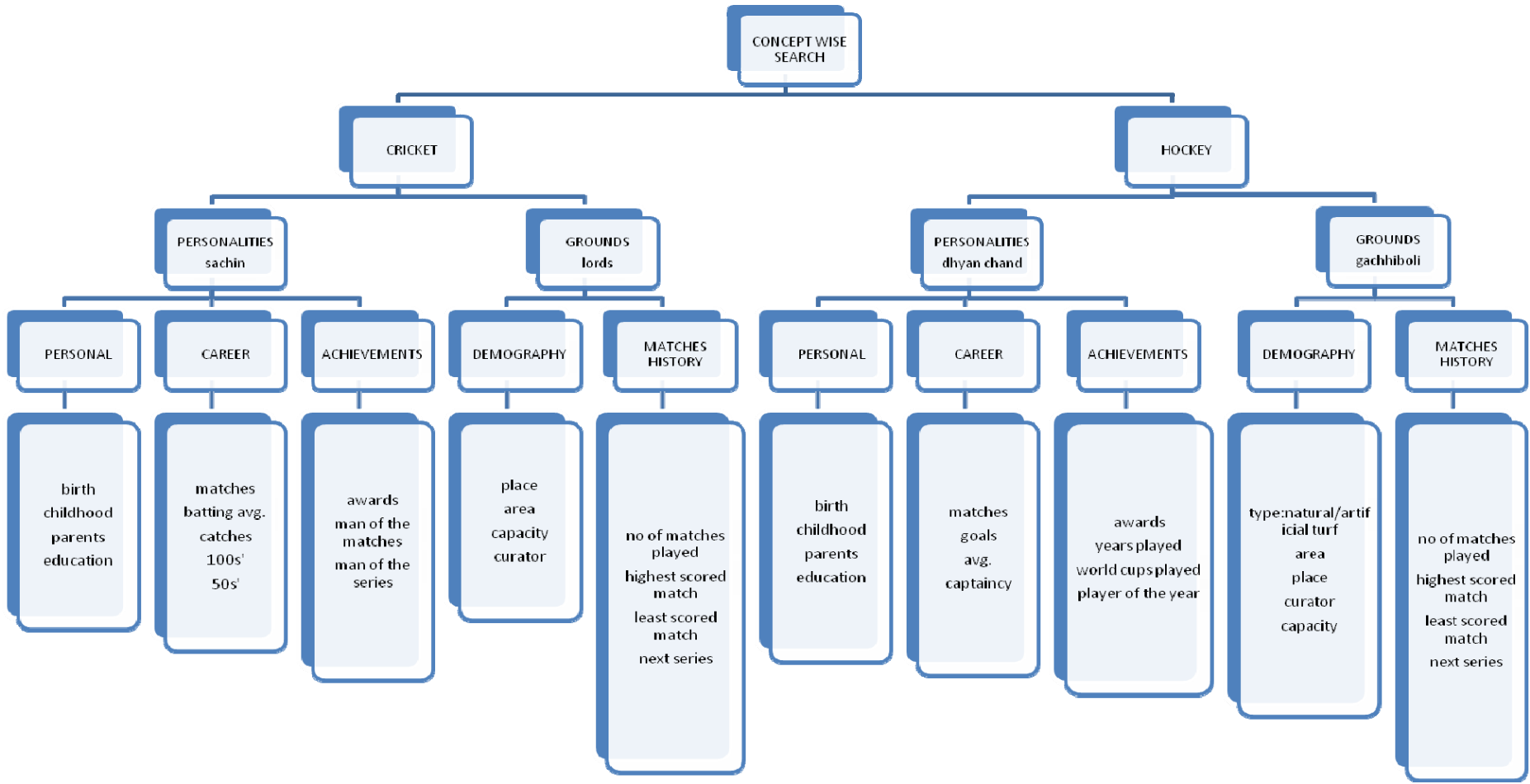

Fig. 2. Tree diagram for extraction of concept based keywords based on the query